\documentclass[12pt,preprint]{aastex}

\shorttitle{Evidence of an asteroid encountering a pulsar}
\shortauthors{Brook et al.}
\begin{document}

\title{Evidence of an asteroid encountering a pulsar}

\author{P. R. Brook\altaffilmark{1} and A. Karastergiou}
\affil{Astrophysics, University of Oxford, Denys Wilkinson Building, Keble Road, Oxford OX1 3RH, UK}
\email{paul.brook@astro.ox.ac.uk}

\author{S. Buchner\altaffilmark{2}}
\affil{Hartebeesthoek Radio Astronomy Observatory, P.O. Box 443, Krugersdorp, 1740, South Africa}

\author{S. J. Roberts}
\affil{Information Engineering, University of Oxford, Parks Road, Oxford OX1 3PJ, UK}

\author{M. J. Keith\altaffilmark{3}, S. Johnston and R.M. Shannon}
\affil{CSIRO Astronomy \& Space Science, Australia Telescope National Facility, P.O. Box 76, Epping, NSW 1710, Australia}

\altaffiltext{1}{CSIRO Astronomy \& Space Science, Australia Telescope National Facility, P.O. Box 76, Epping, NSW 1710, Australia}
\altaffiltext{2}{School of Physics, University of Witwatersrand, Johannesburg, South Africa}
\altaffiltext{3}{Jodrell Bank Centre for Astrophysics, School of Physics and Astronomy, University of Manchester, Manchester M13 9PL, UK}

\begin{abstract}
  Debris disks and asteroid belts are expected to form around young
  pulsars due to fallback material from their original supernova
  explosions. Disk material may migrate inwards and interact with a
  pulsar's magnetosphere, causing changes in torque and emission. Long
  term monitoring of PSR J0738$-$4042 reveals both effects. The pulse
  shape changes multiple times between 1988 and 2012. The torque,
  inferred via the derivative of the rotational period, changes
  abruptly from September 2005. This change is accompanied by an
  emergent radio component that drifts with respect to the rest of the
  pulse. No known intrinsic pulsar processes can explain these timing
  and radio emission signatures. The data lead us to postulate that we
  are witnessing an encounter with an asteroid or in-falling debris
  from a disk.
\end{abstract}


\section{Introduction}

Pulsars are used as unique high-precision clocks in experimental
astrophysics, primarily because of two fundamental characteristics.
First, the average shape of the emitted radio beam (known as the pulse
profile) remains remarkably stable over decades of observation,
despite substantial variability from pulse to pulse.  Second, the high
moment of inertia and fast rotation of neutron stars results in
extreme rotational stability, which can be measured by exploiting the
stable pulse profile.  The rotational frequency $\nu$ of all pulsars
is gradually decreasing due to the loss of energy from magnetic dipole
radiation. The time derivative of the frequency is known as the
spin-down rate $\dot{\nu}$. The stability of pulsars allows timing
models to account for phenomena that affect the arrival times of
pulses on Earth, including relative motions, propagation effects, and
general relativistic effects \citep{2006MNRAS.372.1549E}.

In the last few years, evidence has been emerging that challenges both
of the characteristics above. A small number of pulsars have been
identified \citep{2010Sci...329..408L}, in which the average pulse
profile switches between states on long timescales, some pulsars also
showing a correlated change in $\dot{\nu}$. A similar but extreme
example of variability is observed in a group of intermittent pulsars
which go through a quasi-periodic cycle of switching on and off, on
timescales of months to years \citep{2006Sci...312..549K,
  2012ApJ...746...63C, 2012ApJ...758..141L}.  In these objects, each
of their two states is associated with a different spin-down rate.  In
both cases of state-switching pulsars, the changes in $\dot{\nu}$ and
emission have been attributed to the torque induced on the neutron
star by changing magnetospheric currents \citep{2010MNRAS.408L..41T}.

Neutron star glitches, characterized by a discrete increase and
gradual relaxation of the rotational frequency, have also recently
been linked to pulse profile variability in radio pulsars
\citep{2011MNRAS.411.1917W, 2013MNRAS.432.3080K}. Further examples of
glitches and irregular spin-down properties, associated with emission
variability, have been seen in magnetars \citep{2006csxs.book..547W},
where dramatic profile changes, related to changes in the magnetic
field structure have been observed both in X-rays and radio
\citep[e.g.][]{2007ApJ...663..497C}.

Profile and timing variability reduce the potential to use particular
pulsars for the detection of faint signals from a stochastic
gravitational wave background. Investigating potential causes of
magnetospheric changes is crucial when aiming to correct for timing
variability.

\citet{2008ApJ...682.1152C} describe the process by which asteroids,
formed from supernova fallback material, may enter the magnetosphere
of a pulsar and affect both the pulse profile and $\dot{\nu}$.
Further evidence for planetary and disk systems around neutron stars
is described in \citet{2013ApJ...766....5S} and references
therein. Shannon et al. show that the timing of PSR~B1937+21 is
consistent with the presence of an asteroid belt. They also refer to
other examples of planetary and disk systems around neutron stars,
such as the planets around PSR~B1257+12 \citep{1992Natur.355..145W}
and PSR~B1620-26 \citep{1999ApJ...523..763T}, the dust disk around
magnetar 4U 0142+61 \citep{2006Natur.440..772W}, and the unusual
$\gamma$-ray burst GRB~101225A, thought to be due to a minor body
falling onto a neutron star \citep{2011Natur.480...69C}. Also of
interest is their reference to circumstantial evidence for asteroid
belts around white dwarfs, suggesting that rocky bodies may exist
around post main sequence stars.

PSR J0738$-$4042 is a bright, radio-emitting neutron star with
rotational properties similar to the main population of middle-aged,
isolated, radio pulsars. It has $\nu$ and $\dot{\nu}$ values of approximately 2.667~
s$^{-1}$ and $-1.15\times10^{-14}$~s$^{-2}$ respectively.
This pulsar has been the target of a long-term and
high-cadence monitoring campaign at the Hartebeesthoek Radio Astronomy
Observatory (HartRAO) in South Africa, and a more recent
campaign at the Parkes radio telescope in Australia. The result is a
unique 24-year dataset in which we have discovered a short period of
significant perturbations in $\dot{\nu}$ that coincide with the
appearance of a new component in the average pulse profile, first
noted by \citet{2011MNRAS.415..251K}.  In section 2, we provide
details about the observations of PSR~J0738-4042. In section 3, we
present the emission and timing variability observed.  The data reveal
the abrupt changes in rotational stability and emission properties,
which have previously only been theoretically predicted as
consequences of a pulsar-asteroid encounter
\citep{2008ApJ...682.1152C}, an interpretation which we discuss in
detail in section 4.

\section{Observations}

Data from PSR J0738$-$4042 were collected from September 1988 to
September 2012 using the 26~m antenna of HartRAO. Observations were
made at intervals from 1 to 14 days using receivers at 1600~MHz (1664
and 1668~MHz) or 2300~MHz (2270 and 2273~MHz). The 1600~MHz data have
been used here for the profile stability analysis, whereas both data
sets were used to obtain timing results. The observations were made of
a single polarization: left hand circular. However, the degree of
circular polarization in this pulsar is low
\citep{2011MNRAS.415..251K}, resulting in a negligible difference
between true total power profiles and the HartRAO data.  During this
period the backend provided a single frequency channel of 10 MHz
(until April 2003) and then of 8 MHz at 1600 MHz and of 16 MHz at 2300
MHz. Dispersion due to the interstellar medium is limited to $\sim$
3~ms across a 10 MHz band centred at 1600 MHz.  Observations usually
consisted of three consecutive 15 minute (2400 pulsar period)
integrations, each resulting in a single integrated profile. There is
a gap in the regular coverage from April 1999 - August 2000 (MJD 51295
- 51775) due to equipment upgrade. In order to determine the pulse
time of arrival at each epoch, an analytic profile consisting of three
Gaussian components was fitted to the integrated profiles. The arrival
times were then processed using {\bf TEMPO2}
\citep{2006MNRAS.369..655H}.

Since 2007, PSR J0738$-$4042 has been observed on a roughly monthly
basis at 1369 MHz with the Parkes radio telescope as part of the Fermi
timing programme \citep{2010PASA...27...64W}. The data were recorded
with one of the Parkes Digital Filterbank systems (PDFB1/2/3/4) with a
total bandwidth of 256 MHz in 1024 frequency channels. Observations
are calibrated for differential gain and phase between the feeds using
measurements of a noise diode coupled to the receptors in the feed. To
correct for cross-coupling of the receptors in the Multibeam receiver,
a model of the Jones matrix was used for the receiver computed by
observation of the bright pulsar PSR J0437$-$4715 over the entire
range of hour angles visible, using the measurement equation modelling
technique \citep{2004ApJS..152..129V}.

\section{Variability in PSR J0738-4042}

The radio emission and timing history of PSR J0738$-$4042 between 1988
and 2012 are presented in Figure 1. To emphasize the variability of
the average pulse profile, we have subtracted a constant model profile
from each observation. Panel F shows the residual between the data and
the model as a function of epoch and rotational pulse phase, centred
around the pulse peak and with a temporal resolution of
$\sim$ 1.46~ms. After 2008, the observations have a much higher signal-to-noise ratio (S/N), reflecting
the higher sensitivity of the Parkes telescope. Each profile included in the HartRAO section of panel F is an average
of 8 consecutive observations, in order to increase the S/N.
To obtain profiles evenly spaced in time,
we averaged 8 profiles per 10-day interval. If a given 10-day interval
contained fewer than 8 observations, it was extended by 10 days until
8 profiles could be averaged. This average profile was then used for
both 10-day intervals from which the data were taken. The averaging
process was not repeated for the Parkes data. However, the data were
still divided into 10-day windows, to maintain the same sampling as
for HartRAO. If multiple observations existed in a single window, the
profile with the highest S/N was used. If no observations existed within a
given 10-day window, the profile from the previous window was carried
over. We have divided the data into five intervals with distinct profile
shapes, shown in panels A$-$E. The history of $\dot{\nu}$, computed at
25-day intervals, is plotted in panel G.
\begin{figure}[htp]
\begin{center}
\label{threepanels}
\includegraphics[width=170mm]{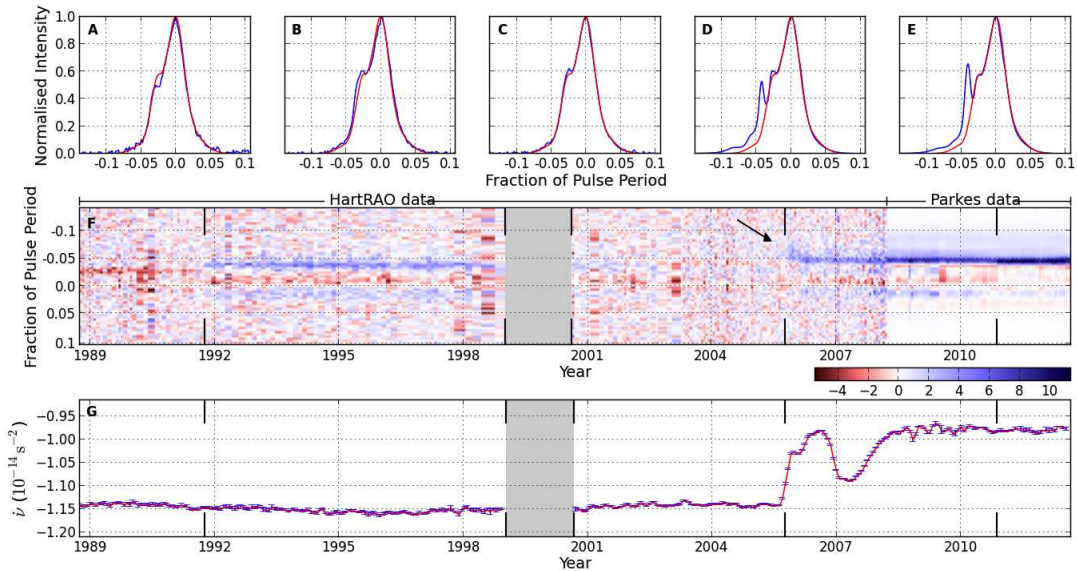}
\caption{Variations in the profile shape and spin-down rate seen in
  PSR J0738$-$4042. Profiles are observed at 1600 MHz with
  HartRAO and at 1369 MHz with the Parkes telescope. Panels A$-$E:
  The blue trace denotes the median pulse profile for each of 5
  intervals over the 24-year dataset, which are demarcated in panels F
  and G. The red trace in each plot is a constant model profile which
  represents the median profile for all of the HartRAO data. Panel F:
  Map showing the difference between data and the constant model, in
  units of the HartRAO off-pulse standard deviation. 
  The epochs at which data were collected from both telescopes were
  used to normalize the Parkes data to the HartRAO scale. The arrow points
  to drifting emission changes which precede the emergence of a new
  persistent profile component.
  Panel G: The pulsar spin-down rate as
  a function of time.}
\end{center}
\end{figure}
A dramatic change in pulse shape occurs simultaneously with an abrupt
change in torque from September 2005. This change is accompanied by a
radio component that initially drifts with respect to the rest of the
pulse, before becoming the stable new profile feature reported by
\citet{2011MNRAS.415..251K}.
The drifting feature and its relationship with $\dot{\nu}$ are shown
in high contrast in Figure 2. Here, $\dot{\nu}$ is interpolated using
a non-parametric model, thus avoiding the requirement to chose an
explicit functional form for the interpolation
\citep{Rasmussen&Williams, Roberts_etal}.  The drift occurs over
$\sim$ 0.02 of the pulse period and has a duration of $\sim$ 100
days. As it begins, a pronounced change in $\dot{\nu}$ is simultaneously
seen in the interpolated curve. The value of $\dot{\nu}$ is relatively stable both
before the 2005 event, at $\sim -1.14 \times 10^{-14}$~s$^{-2}$ and
$\sim$ 1000 days later at $\sim -0.98 \times 10^{-14}$~s$^{-2}$.

For the timing analysis presented here, the pulsar times-of-arrival
were computed by a standard technique of cross-correlating the
observed profile with a template. The template does not include the
new component for the HartRAO data, while it does for Parkes. Values
of $\dot{\nu}$ were then determined from the times-of-arrival
at various observing frequencies: 1600 MHz and 2300 MHz at HartRAO and
1369 MHz at Parkes. The values were calculated using
the {\it glitch} plugin to {\bf TEMPO2}. Overlapping regions of 150
days were selected at 25 day intervals and a timing model of $\nu$ and
$\dot{\nu}$ was fitted within each region. Values of $\dot{\nu}$ with
uncertainties in excess of $10^{-16}$ s$^{-2}$, were considered
unreliable and not included in Figure 1.

\begin{figure}[htp]
\begin{center}
\label{phaseemission}
\includegraphics[width=160mm]{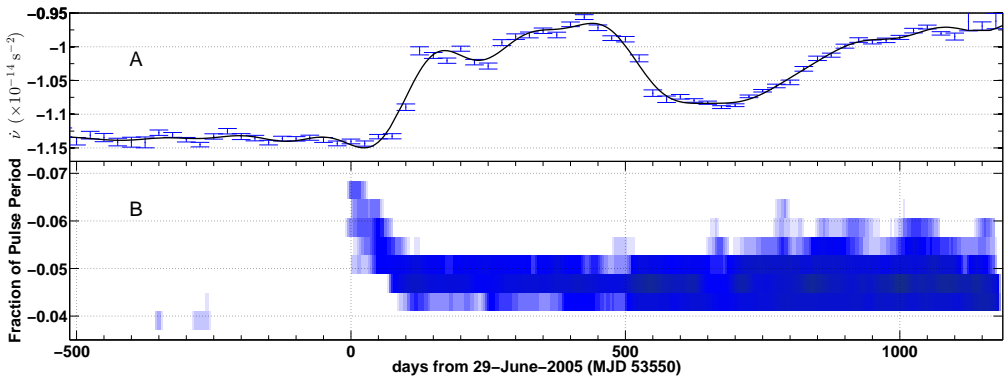}
\caption{Panel A: $\dot{\nu}$ as a function of time, as computed over
  a 1700 day period referenced to 29 June 2005. The curve is
  interpolated from data points on which it is overlaid. Panel B: The
  profile residuals as they appear over the same time period. High
  contrast is used in order to emphasize the drifting feature.}
\end{center}
\end{figure}

\section{Interpretation and discussion}

In considering the source of the variations in magnetospheric
currents, which result in the observed $\dot{\nu}$ changes and
simultaneous changes in pulse profile, the possibility that
PSR~J0738$-$4042 encountered an asteroid or debris from a disk must be
explored. \citet{2008ApJ...682.1152C}, discuss how asteroids, formed
from supernova fallback material, may enter the magnetosphere of a
pulsar as a result of the migration due to collisions, orbital
perturbations and the Yarkovsky effect, or by direct injection from
eccentric orbits. A small body, falling towards a pulsar, is
evaporated and ionized via pulsar radiation.  The remaining charges,
which are electrically captured by gap regions, can perturb particle
acceleration in various ways. When accelerated to relativistic
energies, charges can produce gamma-rays, which give rise to a
pair-production cascade.  A quiescent region of the magnetosphere can
be activated in this way, leading to new observable emission
components. The injected charged particles may also diminish the
electric field of a gap region, consequently attenuating an existing
pair-production cascade.  In the canonical pulsar, radio emission
generated at higher altitudes on a set of dipolar magnetic field lines
will be beamed at larger angles with respect to the magnetic axis than
emission at lower altitudes.  Two interpretations of the drift shown
in Figure 2 are either azimuthal movement of an emitting region at a
given height, or an emission region moving in height along particular
magnetic field lines. In the context of the former, the low rate of
phase drift does not correspond to any process known, or seen
previously in other pulsars \citep{2006A&A...445..243W}. Additionally,
Cordes \& Shannon note that a change in pair production can result in
a change in emission altitude for a given frequency, due to the plasma
frequency dependence on height. We hypothesize hereafter that the
phase drift is attributed to a decrease in emission height.

In the case of a dipolar magnetic field emitting over multiple
heights, there is a relationship between the angular radius of the
field lines $\rho$ at a given height and the observed pulse phase of
the emission $\phi$; the lower the altitude, the closer the emission
component will be to the centre of the profile. This relationship can
be derived using:
\begin{equation}
\sin^{2} \left(\frac{W}{4}\right) = \frac{\sin^{2}(\rho/2) - \sin^{2}(\beta/2)}{\sin\alpha \sin(\alpha + \beta)},
\end{equation}
where $\alpha$ is the angle of the magnetic axis with respect to the
rotation axis, $\beta$ is the closest approach of the line of sight to
the magnetic axis and W is the total width of the pulse profile
\citep{1984A&A...132..312G}. The observed pulse phase $\phi$, measured
from the peak of the profile, can be substituted directly for W/2. The
values of $\phi$ at which the drift begins and ends are measured in
degrees to be $\sim 23.4^{\circ}$ and $\sim 16.2^{\circ}$
respectively. In order to obtain a value for $\rho$, at the begining
and end of the drift ($\rho_{begin}$ and $\rho_{end}$), $\alpha$ and
$\beta$ for the pulsar are needed. A value for $\beta$ is obtained via
the following equation:
\begin{equation}
\beta = \sin^{-1} \left(\frac{\sin~\alpha}{|d\chi/d\phi|_{max}}\right),
\end{equation}
where $|d\chi/d\phi|_{max}$ is the maximum rate of change of the
polarization position angle occuring around the centre of the pulse
profile \citep{1993ApJ...405..285R}. For PSR J0738$-$4042,
$|d\chi/d\phi|_{max}$ is $\sim 3$ \citep{2011MNRAS.415..251K}.  The
value for $\alpha$ is not easily constrained.
Through calculations for various possible values of $\alpha$, we find
that $\rho_{begin}/\rho_{end}$ is largely independent of $\alpha$. For
emission originating close to the magnetic axis, $\rho$ is related to
the height $H$ in the following way \citep{2007MNRAS.380.1678K}:
\begin{equation}
\rho \sim \sqrt{\frac{9\pi H}{2cP}}
\end{equation}
and therefore,
\begin{equation}
  \frac{H_{begin}}{H_{end}} =
  \left(\frac{\rho_{begin}}{\rho_{end}}\right)^{2}.
\end{equation}
As $\rho_{begin}/\rho_{end}$ is insensitive to $\alpha$, so too is
$H_{begin}/H_{end}$ which has a value of $\sim$ 1.5. The conclusion,
therefore, is that the change in height of the drifting emission
region is around half of its final emission height.

A reconfiguration in current within a pulsar magnetosphere would
simultaneously affect the braking torque and, therefore, the spin-down
rate of the pulsar.  The 2005 change in $\dot{\nu}$ can be interpreted
as a reduction in the total outflowing plasma above the polar
caps. The magnitude of the current change, can be inferred from the
change in $\dot{\nu}$ following \citet{2006Sci...312..549K}.  The
difference between the two extreme $\dot{\nu}$ values corresponds to a
reduction in the charge density $\rho$ of $\sim 7 \times 10^{-9} $ C cm$^{-3}$,
where $\rho = 3I\Delta\dot{\nu}/R_{pc}^{4}B_{0}$, the moment of
inertia $I$ is taken to be 10$^{45}$ g cm$^{2}$, the magnetic field
$B_{0} = 3.2 \times 10^{19}\sqrt{-\dot{\nu}/\nu^{3}}$ Gauss, polar cap
radius $R_{pc} = \sqrt{2\pi R^{3}\nu/c}$ and where the neutron star
radius $R$ is taken to be $10^{6}$ cm. We can relate the difference in
charge density associated with the two spin-down states to mass
supplied to the pulsar, by multiplying it by the speed of light, the
polar cap area and the duration of the new spin-down state. Between
2005 and today, this amounts to $\sim 10^{15}$~g, which lies within
the range of known solar system asteroid masses, and is consistent
with the mass range of asteroids around neutron stars proposed by
\citet{2008ApJ...682.1152C}.  This line of reasoning provides us with
a testable hypothesis; at the point in time when the injected mass is
exhausted, we would expect the pulsar to return to its previous
spin-down state.

If we are witnessing an encounter between the pulsar and an asteroid,
the question arises as to whether and why an event would be
unique. Although the largest emission and $\dot{\nu}$ change in PSR
J0738$-$4042 occur in 2005, Figure 1 shows a similar but less
pronounced emission increase in 1992 along the leading edge of the
pulsar. Figure 1 also shows that in 2010, the new component
experiences significant and sudden growth, seen clearly in panels D
and E. Neither the 1992 or the 2010 emission changes are accompanied
by significant $\dot{\nu}$ changes. We also note the multiple profile
modes of PSR J0738$-$4042, as opposed to the seemingly bimodal nature
of the state-switching pulsars in \citet{2010Sci...329..408L}.

The 1992 and 2010 emission changes could also be caused by material
entering the magnetosphere, but the smaller effect on pulse profile
and the apparent stability of $\dot{\nu}$ during the emission changes
may suggest smaller amounts of in-falling matter. The profile changes
could be due to a large orbiting body, such as a planet, which
periodically disrupts debris in the fallback disk and precipitates
inward migration. As an initial test of this hypothesis, we have
simulated the perturbation to timing measurements of this pulsar by an
orbiting planet. We performed a periodicity analysis of the residuals
in 1000 day segments, using the Cholesky pre-whitening method of
\citet{2011MNRAS.418..561C} and fitting for sinusoids with frequencies
linearly spaced from 0.001 per day to 10 per day. We then computed a
weighted mean of the resulting periodograms. Using a 5$\sigma$
threshold we found no significant signals at long periods, and only
two significant periodic signals, at exactly 1 per day and 2 per day,
the origin of which is unclear. By injecting a simulated planetary
signature with a circular orbit of radius $10^{9}$~m (typical
gravitational tidal radius \citep{2008ApJ...682.1152C}) we rule out
any such planet with mass greater than 6$\times 10^{28}$~g. A planet
of smaller mass may, therefore, exist without imprinting its signature
on the timing data of this pulsar. As a future step, we intend to test
the idea of periodic disruption of a debris disk by a planet, by
monitoring PSR J0738$-$4042 for years to come, to see whether the
emission profile and $\dot{\nu}$ will periodically vary.

The investigation of alternative interpretations, and glitches in particular,
forms part of our future plans. The observed emission variability
and rotational instabilities of magnetars are of interest in this
context, especially given the latest estimates of the surface magnetic
field of SGR 0418+5729 \citep{2013ApJ...770...65R}, which provides a
link between the high magnetic field neutron stars and radio
pulsars. Variability may prove to be another common characteristic
between these populations of neutron stars. 

The Parkes radio telescope is part of the Australia Telescope National
Facility which is funded by the Commonwealth of Australia for
operation as a National Facility managed by CSIRO. Data taken at Parkes
is available via a public archive. Consult the web page
http://www.atnf.csiro.au/research/pulsar/index.php?n=Main.ANDSATNF
for details. P.R.B. is grateful to
the Science and Technology Facilities Council and CSIRO Astronomy and
Space Science for support throughout this work. We thank the referee Scott Ransom
for valuable comments that helped to improve the text.\\

\end{document}